# A Survey: *Various Techniques of Image Compression*

Gaurav Vijayvargiya
Dept. of CSE
UIT- RGPV
Bhopal, India
vijaygaurav1212@gmail.com

Dr. Sanjay Silakari
Dept. of CSE
UIT- RGPV
Bhopal, India
ssilakari@yahoo.com

Dr.Rajeev Pandey
Dept. of CSE
UIT- RGPV
Bhopal, India
rajeev98iet@gmail.com

*Abstract*—**This paper addresses about various image compression techniques. On the basis of analyzing the various image compression techniques this paper presents a survey of existing research papers. In this paper we analyze different types of existing method of image compression. Compression of an image is significantly different then compression of binary raw data. To solve these use different types of techniques for image compression. Now there is question may be arise that how to image compress and which types of technique is used. For this purpose there are basically two types are method are introduced namely lossless and lossy image compression techniques. In present time some other techniques are added with basic method. In some area neural network genetic algorithms are used for image compression.**

## I. INTRODUCTION

An image is an artifact that depicts or records visual perception. Images are important documents today; to work with them in some applications there is need to be compressed. Compression is more or less it depends on our aim of the application. Image compression plays a very important role in the transmission and storage of image data as a result of and storage limitations. The main aim of image compression is to represent an image in the fewest number of bits without losing the essential information content within an original image. Compression [3] techniques are being rapidly developed for compress large data files such as images. With the increasing growth of technology a huge amount of image data must be handled to be stored in a proper way using efficient techniques usually succeed in compressing images. There are some algorithms that perform this compression in different ways; some are lossless and lossy. Lossless keep the same information as the original image and in lossy some information loss when compressing the image. Some of these compression techniques are designed for the specific kinds of images, so they will not be so good for other kinds of images. In Some algorithms let us change few parameters they use to adjust the compression better to the image. Image compression is an application of data compression that encodes the original image with fewer bits. The objective of image compression [1] is to reduce the redundancy of the image and to store or transmit data in an efficient form.
The compression ratio is defined as follows:

$$C_r = N1/N2$$

where N1 is the data of the actual image and N2 is the data of compressed image.

## II. IMAGE COMPRESSION

Image compression addresses the problem of reducing the amount of information required to represent a digital image. It is a process intended to yield a compact representation of an image, thereby reducing the image storage transmission requirements. Every image will have redundant data. Redundancy means the duplication of data in the image. Either it may be repeating pixel across the image or pattern, which is repeated more frequently in the image.The image compression occurs by taking benefit of redundant information of in the image. Reduction of redundancy provides helps to achieve a saving of storage space of an image. Image compression is achieved when one or more of these redundancies are reduced or eliminated. In image compression, three basic data redundancies can be identified and exploited. Compression is achieved by the removal of one or more of the three basic data redundancies.

### A. Inter Pixel Redundancy

In image neighbouring pixels are not statistically independent. It is due to the correlation between the neighboring pixels of an image. This type of redundancy is called Inter-pixel redundancy. This type of redundancy is sometime also called spatial redundancy. This redundancy can be explored in several ways, one of which is by predicting a pixel value based on the values of its neighboring pixels. In order to do so, the original 2-D array of pixels is usually mapped into a different format, e.g., an array of differences between adjacent pixels. If the original image [20] pixels can be reconstructed from the transformed data set the mapping is said to be reversible.

## B. Coding Redundancy

Consists in using variable length code words selected as to match the statistics of the original source, in this case, the image itself or a processed version of its pixel values. This type of coding is always reversible and usually implemented using lookup tables (LUTs). Examples of image coding schemes that explore coding redundancy are the Huffman codes and the arithmetic coding technique.

## C. Psycho Visual Redundancy

Many experiments on the psycho physical aspects of human vision have proven that the human eye does not respond with equal sensitivity to all incoming visual information; some pieces of information are more important than others. Most of the image coding algorithms in use today exploit this type of redundancy, such as the Discrete Cosine Transform (DCT) based algorithm at the heart of the JPEG encoding standard.

## III. BENEFITS OF COMPRESSION

- It provides a believable cost savings involved with sending less data over the switched telephone network where the cost of the call is really usually based upon its duration.
- It not only reduces storage requirements but also overall execution time.
- It reduces the probability of transmission errors since fewer bits are transferred.
- It provides a level of security against unlawful monitoring.

## IV. COMPARISON BETWEEN LOSSLESS AND LOSSY TECHNIQUES

In lossless compression schemes, the reconstructed image, after compression, is numerically identical to the original image. However lossless compression can only a achieve a modest amount of compression. An image reconstructed following lossy compression contains degradation relative to the original. Often this is because the compression scheme completely discards redundant information. However, lossy schemes are capable of achieving much higher compression.

## A. Types of Image Compression

On the bases of our requirements image compression techniques are broadly bifurcated in following two major categories.
- Lossless image compression
- Lossy image compression

### 1) Lossless Compression Techniques:

Lossless compression compresses the image by encoding all the information from the original file, so when the image is decompressed, it will be exactly identical to the original image. Examples of lossless [2] image compression are PNG and GIF. When to use a certain image compression format really depends on what is being compressed.

### a) Run Length Encoding:

Run-length encoding (RLE) is a very simple form of image compression in which runs of data are stored as a single data value and count, rather than as the original run. It is used for sequential [19] data and it is helpful for repetitive data. In this technique replaces sequences of identical symbol (pixel), called runs. The Run length code for a grayscale image is represented by a sequence { $V_i$, $R_i$ } where $V_i$ is the intensity of pixel and $R_i$ refers to the number of consecutive pixels with the intensity $V_i$ as shown in the figure. This is most useful on data that contains many such runs for example, simple graphic images such as icons, line drawings, and animations. It is not useful with files that don't have many runs as it could greatly increase the file size. Run-length encoding performs lossless image compression. Run-length encoding is used in fax machines.

| 65 | 65 | 65 | 70 | 70 | 70 | 70 | 72 | 72 | 72 |
|----|----|----|----|----|----|----|----|----|----|

| {65,3} | {70,4} | {72,3} |
|--------|--------|--------|

### b) Entropy Encoding:

In information theory an entropy encoding is a lossless data compression scheme that is independent of the specific characteristics of the medium. One of the main types of entropy coding creates and assigns a unique prefix-free code for each unique symbol that occurs in the input. These entropy encoders then compress the image by replacing each fixed-length input symbol with the corresponding variable-length prefix free output codeword.

### c) Huffman Encoding:

In computer science and information theory, Huffman coding is an entropy encoding algorithm used for lossless data compression. It was developed by Huffman. Huffman coding [8] today is often used as a "back-end" to some other compression methods. The term refers to the use of a variable-length code table for encoding a source symbol where the variable-length code table has been derived in a particular way based on the estimated probability of occurrence for each possible value of the source symbol. The pixels in the image are treated as symbols. The symbols which occur more frequently are assigned a smaller number of bits, while the symbols that occur less frequently are assigned a relatively larger number of bits. Huffman code is a prefix code. This means that the (binary) code of any symbol is not the prefix of the code of any other symbol.

### d) Arithmetic Coding :

Arithmetic coding is a form of entropy encoding used in lossless data compression. Normally, a string of characters such as the words "hello there" is represented using a fixed number of bits per character, as in the ASCII code. When a string is converted to arithmetic encoding, frequently used characters will be stored with little bits and not-so-frequently occurring characters will be stored with more bits, resulting in fewer bits used in total. Arithmetic coding differs from other forms of entropy encoding such as Huffman coding [10] in that rather than separating the input into component symbols and replacing each with a code, arithmetic coding encodes the entire message into a single number.

*e) Lempel–Ziv–Welch Coding:* Lempel–Ziv–Welch (LZW) is a universal lossless data compression algorithm created by Abraham Lempel, Jacob Ziv, and Terry Welch. It was published by Welch in 1984 as an improved implementation of the LZ78 algorithm published by Lempel and Ziv in 1978. LZW is a dictionary based coding. Dictionary based coding can be static or dynamic. In static dictionary coding, dictionary is fixed when the encoding and decoding processes. In dynamic dictionary coding, dictionary is updated on fly. The algorithm is simple to implement, and has the potential for very high throughput in hardware implementations. It was the algorithm of the widely used UNIX file compression utility compress, and is used in the GIF image format. LZW compression became the first widely used universal image compression method on computers. A large English text file can typically be compressed via LZW to about half its original size.

*2) Lossy Compression Techniques:*
Lossy compression as the name implies leads to loss of some information. The compressed image is similar to the original uncompressed image but not just like the previous as in the process of compression [9] some information concerning the image has been lost. They are typically suited to images. The most common example of lossy compression is JPEG. An algorithm that restores the presentation to be the same as the original image are known as lossy techniques. Reconstruction of the image is an approximation of the original image, therefore the need of measuring of the quality of the image for lossy compression technique. Lossy compression technique provides a higher compression ratio than lossless compression.

Major performance considerations of a lossy compression scheme include:
- Compression ratio
- Signal to noise ratio
- Speed of encoding & decoding

Lossy image compression techniques include following schemes:
*a) Scalar Quantization:* The most common type of quantization is known as scalar quantization. Scalar quantization, typically denoted as $Y=Q(x)$, is the process of using a quantization function $Q$ to map a scalar (one-dimensional) input value x to a scalar output value Y. Scalar quantization can be as simple and intuitive as rounding high-precision numbers to the nearest integer, or to the nearest multiple of some other unit of precision.

*b) Vector Quantization:* Vector quantization (VQ) is a classical quantization technique from signal processing which allows the modeling of probability density functions by the distribution of prototype vectors. It was originally used for image compression. It works by dividing a large set of points (vectors) into groups having approximately the same number of points closest to them. The density matching property of vector quantization is powerful, especially for identifying the density of large and high-dimensioned data. Since data points are represented by the index of their closest centroid, commonly occurring data have low error, and rare data high error. This is why VQ is suitable for lossy data compression. It can also be used for lossy data correction and density estimation.

## V. LITERATURE SURVEY

In 2010, Jau-Ji Shen et al presents vector quantization based image compression technique [5]. In this paper they adjust the encoding of the difference map between the original image and after that it's restored in VQ compressed version. Its experimental results show that although there scheme needs to provide extra data, it can substantially improve the quality of VQ compressed images, and further be adjusted depending on the difference map from the lossy compression to lossless compression.

In 2011, Suresh Yerva, et al presents the approach of the lossless image compression using the novel concept of image [6] folding. In this proposed method uses the property of adjacent neighbor redundancy for the prediction. In this method, column folding followed by row folding is applied iteratively on the image till the image size reduces to a smaller pre-defined value. The proposed method is compared with the existing standard lossless image compression algorithms and the results show comparative performance. Data folding technique is a simple approach for compression that provides good compression efficiency and has lower computational complexity as compared to the standard SPIHT technique for lossless compression.

In 2012, Firas A. Jassim, et al presents a novel method for image compression which is called five module method (FMM). In this method converting each pixel value in 8x8 blocks [7] into a multiple of 5 for each of RGB array. After that the value could be divided by 5 to get new values which are bit length for each pixel and it is less in storage space than the original values which is 8 bits. This paper demonstrates the potential of the FMM based image compression techniques. The advantage of their method is it provided high PSNR (peak signal to noise ratio) although it is low CR (compression ratio). This method is appropriate for bi-level like black and white medical images where the pixel in such images is presented by one byte (8 bit). As a recommendation, a variable module method (X) MM, where X can be any number, may be constructed in latter research.

In 2012, Ashutosh Dwivedi, et al presents a novel hybrid image compression technique. This technique inherits the properties of localizing the global spatial and frequency correlation from wavelets and classification and function approximation tasks from modified forward-only counter propagation neural network (MFOCPN) for image compression. In this scheme several tests are used to investigate the usefulness of the proposed scheme. In this paper, they explore the use of MFO-CPN [11] networks to predict wavelet coefficients for image compression. In this method, they combined the classical wavelet based method with MFO-CPN. The performance of the proposed network is tested for three discrete wavelet transform functions. In this they analysis that Haar wavelet results in higher compression ratio but the quality of the reconstructed image is not good. On the other hand db6 with the same number of wavelet coefficients leads to higher compression ratio with good quality. Overall they found that the application of db6 wavelet in image compression out performs other two.

In 2012, Yi-Fei Tan, et al presents image compression technique based on utilizing reference points coding with threshold values. This paper intends to bring forward an image compression method which is capable to perform both lossy and lossless compression. A threshold [12] value is associated in the compression process, different compression ratios can be achieved by varying the threshold values and lossless compression is performed if the threshold value is set to zero. The proposed method allows the quality of the decompressed image to be determined during the compression process. In this method If the threshold value of a parameter in the proposed method is set to 0, then lossless compression is performed. Lossy compression is achieved when the threshold value of a parameter assumes positive values. Further study can be performed to calculate the optimal threshold value T that should be used.

In 2012, S.Sahami, et al presents a bi-level image compression techniques using neural networks". It is the lossy image compression technique. In this method, the locations of pixels of the image are applied to the input of a multilayer perceptron neural network [13]. The output the network denotes the pixel intensity 0 or 1. The final weights of the trained neural-network are quantized, represented by few bites, Huffman encoded and then stored as the compressed image. Huffman encoded and then stored as the compressed image. In the decompression phase, by applying the pixel locations to the trained network, the output determines the intensity. The results of experiments on more than 4000 different images indicate higher compression rate of the proposed structure compared with the commonly used methods such as comite consultatif international telephonique of telegraphique graphique (CCITT) G4 and joint bi-level image expert group (JBIG2) standards. The results of this technique provide High compression ratios as well as high PSNRs were obtained using the proposed method. In the future they will use activity, pattern based criteria and some complexity measures to adaptively obtain high compression rate.

In 2013, C. Rengarajaswamy, et al presents a novel technique in which done encryption and compression of an image. In this method stream cipher is used for encryption of an image after that SPIHT [14] is used for image compression. In this paper stream cipher encryption is carried out to provide better encryption used. SPIHT compression provides better compression as the size of the larger images can be chosen and can be decompressed with the minimal or no loss in the original image. Thus high and confidential encryption and the best compression rate has been energized to provide better security the main scope or aspiration of this paper is achieved.

In 2013, S. Srikanth, et al presents a technique for image compression which is use different embedded Wavelet based image coding with Huffman-encoder for further compression. In this paper they implemented the SPIHT and EZW algorithms with Huffman encoding [15] using different wavelet families and after that compare the PSNRs and bit rates of these families. These algorithms were tested on different images, and it is seen that the results obtained by these algorithms have good quality and it provides high compression ratio as compared to the previous exist lossless image compression techniques.

In 2013, Pralhadrao V Shantagiri, et al presents a new spatial domain of lossless image compression algorithm for synthetic color image of 24 bits. This proposed algorithm use reduction of size of pixels for the compression of an image. In this the size of pixels [16] is reduced by representing pixel using the only required number of bits instead of 8 bits per color. This proposed algorithm has been applied on asset of test images and the result obtained after applying algorithm is encouraging. In this paper they also compared to Huffman, TIFF, PPM-tree, and GPPM. In this paper, they introduce the principles of PSR (Pixel Size Reduction) lossless image compression algorithm. They also had shows the procedures of compression and decompression of their proposed algorithm. Future work of this paper uses the other tree based lossless image compression algorithm.

In 2013, K. Rajkumar, et al presents an implementation of multiwavelet transform coding for lossless image compression. In this paper the performance of the IMWT (Integer Multiwavelet Transform) for lossless studied. The IMWT provides good result with the image reconstructed. In this paper the performance of the IMWT [17] for lossless compression of images with magnitude set coding have been obtained. In this proposed technique the transform coefficient is coded with a magnitude set of coding & run length encoding technique. The performance of the integer multiwavelet transform for the lossless compression of images was analyzed. It was found that the IMWT can be used for the lossless image compression. The bit rate obtained using the MS-VLI (Magnitude Set-Variable Length Integer Representation) with RLE scheme is about 2.1 bpp (bits per pixel) to 3.1 bpp less then that obtain using MS-VLI without RLE scheme.

In 2013 S. Dharanidharan, et al presents a new modified international data encryption algorithm to encrypt the full image in an efficient secure manner, and encryption after the original file will be segmented and converted to other image file. By using Huffman algorithm the segmented image files are merged and they merge the entire segmented image to compress into a single image. Finally they retrieve a fully decrypted image. Next they find an efficient way to transfer the encrypted images to multipath routing techniques. The above compressed image has been sent to the single pathway and now they enhanced with the multipath routing algorithm, finally they get an efficient transmission and reliable, efficient image.

## VI. CONCLUSION

This paper presents various techniques of image compression. These are still a challenging task for the researchers and academicians. There are mainly two types of image compression techniques exist. Comparing the performance of compression technique is difficult unless identical data sets and performance measures are used. Some of these techniques are obtained good for certain applications like security technologies. After study of all techniques it is found that lossless image compression techniques are most effective over the lossy compression techniques. Lossy provides a higher compression ratio than lossless.

### AUTHORS PROFILE

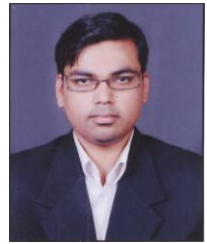

Gaurav Vijayvargiya received his Bachelor's degree in Computer Science & Engineering from BIST-Bhopal, India in 2010.

At present he is pursuing his M.E. degree in Computer Science & Engineering from UIT-RGPV, Bhopal, M.P. India. His research areas are Image Processing, Image Compression, and Image Authentication.

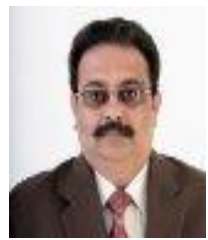

Dr.Sanjay Silakari received his Bachelor's degree in Computer Science & Engineering from SATI, Vidisha in 1991. M.E. (Computer Science & Engineering) from DAVV, Indore in 1998. Ph.D. (Computer Science & Engineering) in 2006 from B.U. Bhopal, M.P. India. He is a member of IEEE.
At present, he is working as Prof. & Head in UIT-RGPV, Bhopal since 2007.

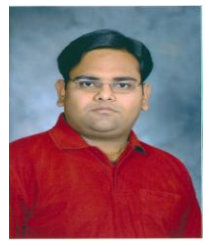

Dr. Rajeev Pandey received his Bachelor's degree in Computer Science & Engineering from IET, DR. B.R.A. University Agra, U.P. India.M.E. (Computer Science & Engineering) from Dr. B.R.A. University, Agra in 2004. Ph.D. in 2010 from Dr.B.R.A. University, Agra, U.P. India. He is also Pursuing Ph.D. (Computer Science & Engineering) from RGPV, Bhopal, M.P. India.
At present, he is working as an Assistant Prof. in UIT-RGPV, Bhopal since 2007.